\documentclass[12pt,epsf]{article}
\usepackage{graphicx, amsmath}

\begin{document}

\sloppy
\begin{flushright}{SIT-HEP/TM-32}
\end{flushright}
\vskip 1.5 truecm
\centerline{\large{\bf Brane inflation without slow-roll}}
\vskip .75 truecm
\centerline{\bf Tomohiro Matsuda
\footnote{matsuda@sit.ac.jp}}
\vskip .4 truecm
\centerline {\it Laboratory of Physics, Saitama Institute of
 Technology,}
\centerline {\it Fusaiji, Okabe-machi, Saitama 369-0293, 
Japan}
\vskip 1. truecm
\makeatletter
\@addtoreset{equation}{section}
\def\theequation{\thesection.\arabic{equation}}
\makeatother
\vskip 1. truecm

\begin{abstract}
\hspace*{\parindent}
The scenario of brane inflation without using the conventional slow-roll
approximations has been investigated. 
Based on the mechanism of generating the curvature perturbations at the
end of inflation, a new brane inflation paradigm was developed. 
The conditions for making a sufficiently large enough number of e-foldings
 and for generating the curvature perturbations without producing
 dangerous relics were also examined. 
Benefits of our scenario are subsequently discussed in detail.  
\end{abstract}

\newpage
\section{Introduction}
\hspace*{\parindent}
In the standard scenario of the inflationary Universe, the observed
density perturbations are produced by a light inflaton that rolls down
its potential. When inflation ends, the inflaton oscillates about
its potential minimum and decays to reheat the Universe. Adiabatic
density perturbations are generated because the scale-invariant
fluctuations of the light inflaton field are different in different
patches. On the other hand, we know that in supersymmetric and
superstring theories there is a serious problem called the 
``$\eta$-problem''.
Without the symmetry that protects the flatness of the inflaton field, a
mass of the order of the Hubble constant will inevitably appear in the
inflaton field, ruining the conventional slow-roll approximation, which
has the typical parameter denoted by ``$\eta$''.
Although it may be possible to construct some inflationary scenarios
where the flatness of the inflaton potential is protected by symmetry,
it is not straightforward to find a situation where the symmetry appears
naturally and all the required conditions for inflation are satisfied
without any fine-tuning.\footnote{It would be very fascinating if
slow-roll inflation could be embedded in MSSM\cite{MSSM-infla}. }
 Thus, we considered an alternative in this paper.
A new inflationary paradigm is developed where the
conventional slow-roll picture does not play an essential role either in
obtaining a large number of e-foldings nor in generating the curvature
perturbations. If such a new scenario is successful, it will perform
even in cases where the inflaton mass is corrected to the order of the
Hubble constant. To be more precise, we will consider a scenario where a
``light field'' is not identified with the inflaton. The most obvious
example in this direction would be the curvaton models\cite{curvaton_1,
curvaton_2}. 
In the curvaton models, the origin of the large-scale curvature
perturbations in the Universe is the late-decay of a massive scalar
field that is called the "curvaton". The curvaton is assumed to be light
during a period of cosmological inflation such that it acquires
scale-invariant fluctuations with the required spectrum. Then, after
inflation, the curvaton starts to oscillate within a radiation
background. The energy density of the curvaton experiences a growth
pattern during this period. The density of the curvaton finally becomes
the dominant part of the total density of the Universe, accounting for
the cosmological curvature perturbations when it decays. The
curvaton paradigm has attracted quite a bit of attention because it was
thought to have obvious advantages. 
For example, since the curvaton is
independent of the inflaton field, there was a
hope\cite{curvaton_liberate} that the curvaton scenario, especially in
models with a low inflationary scale, could cure serious fine-tunings of
the inflation models.\footnote{Many attempts has been made in this
direction. See Ref. \cite{low_inflation} as an example, of which a
complete list and discussion is beyond the scope of this
paper. Cosmological defects may play an essential role in low-scale
brane inflationary scenarios\cite{matsuda_defectinfla}.} 
However, Lyth suggested in his paper\cite{Lyth_constraint} that 
there is a strong bound for the Hubble parameter during inflation.
The bound obtained in Ref. \cite{Lyth_constraint} was a critical
parameter in the inflationary model with a low inflation scale. It was
later suggested in Ref. \cite{matsuda_curvaton, curvaton_added} that the
difficulty could be avoided if an additional inflationary expansion or a
phase transition was present.

More recently, it has been suggested by Lyth\cite{delta-N-Lyth} that
density perturbations can be generated ``at the end of inflation'' by
the number of e-folding fluctuations $\delta N$ induced by a light
scalar field.\footnote{See also Ref. \cite{alternate}, where 
different models generating a contribution to the curvature perturbation
were proposed.}
This novel mechanism is quite simple but very useful.
One does not have to put severe
conditions on the inflaton field, because the
generation of the curvature perturbations is now due to the
fluctuations of a light field, which is independent of the details of the
inflaton dynamic details.
The new mechanism is similar to the curvaton models on this
point. 
However, there is an important advantage in the new mechanism;
unlike the curvatons, one does not have to worry about the serious
conditions that come from the requirement of the successful late-time
dominance and reheating. 

Using this idea, we studied in Ref. \cite{matsuda_elliptic} 
the generation of the curvature perturbations without using slow-roll
approximations. We considered the fluctuations appeared on the
equipotential surface of the multi-field potential. The condition
considered in Ref. \cite{matsuda_elliptic} is very common in brane
inflationary models when extra dimensions are added.\footnote{In
Ref. \cite{Riotto-Lyth}, one can find another 
useful discussions on this point.}
The benefit of the model is that concern over symmetry is not warranted,
as the potential along the equipotential surface is flat by
definition. We also considered a brane inflationary model where a light
field appears ``at a distance'' from the moving brane. This idea fits in
well with the generic requirements of the KKLT model.
To be more precise, we demonstrated that the conventional $\eta$-problem
related to the inflationary brane position is not a serious problem if
there is a symmetry enhancement ``at the tip'' of the throat.
This scenario is very natural in the brane Universe.

Although the generation of the density perturbations was
successful in Ref. \cite{matsuda_elliptic}, it is still ambiguous if one 
can achieve a large number of e-foldings without using the usual
slow-roll approximations. 
Dangerous relics in this type of model should also be
examined, since the isometry at the tip makes the Kaluza-Klein mode with
the angular momentum very sustainable. Considering the above conditions,
we studied a new paradigm of brane inflation, which is not restricted by
the conventional slow-roll conditions nor by the curvaton conditions. We
will show that our settings are very natural in the brane Universe and
can successfully generate the required number of e-foldings, curvature
perturbations and relic densities. We will consider a model where
thermal inflation is followed by fast-roll inflation. In this model, the
hybrid-type potential plays a critical role in both inflationary
epochs. A moving brane is attracted to a distant brane during thermal
inflation, and is subsequently detached at a low temperature.
The moving brane rolls down the potential toward the point where
inflation ends with a brane collision. Although thermal inflation is not
essential in generating the curvature perturbations, it fixes the
initial conditions and supports the succeeding stage of the
fast-roll\cite{fast-roll} or DBI\cite{Tong-DBI} inflation. 
Unlike the conventional scenario of thermal inflation followed by
fast-roll inflation, where massless modes play a significant role to 
generate the unwanted peaks in the spectrum, no such massless modes
appear in our scenario.
We will further discuss this point in Sect.\ref{Thermal-section}.

In our scenario, it is better to lower the scale of inflation so that
one can obtain a large number of e-foldings.\footnote{We will derive this 
condition in Sec.\ref{Thermal-section} by using the idea of fast-roll
inflation. 
Alternatively, one may use the idea of DBI inflatio\cite{Tong-DBI} to
obtain a large enough number of e-foldings.
In the latter case, however, ``slow-roll'' parameters appear in the
theory.
Since we are considering brane inflationary models where ``slow-roll''
 is not essential, it should be unfair to claim that the problem of the
 number of e-foldings is
 solved by using the DBI inflationary models.}
On the other hand, as we will discuss in Sect.\ref{Curva-Densi},
the requirements from the generation of the curvature
perturbations places a lower bound on the inflationary scale, which
cannot be satisfied by the condition obtained from e-foldings, if no
fine-tunings are made. This problem can be solved if there is a
preceding short period of inflationary expansion; the solution of which
is similar to the mechanism discussed for the curvatons in
Ref. \cite{matsuda_curvaton}. 
We will discuss this mechanism in Sect.\ref{Curva-Densi}.
Aside from the curvature perturbation generation, one might think that
the dangerous overproduction of unwanted gravitational relics might be
critical, since the enhanced isometries are known to make the
Kaluza-Klein modes completely stable. It was discussed in
Ref. \cite{KK-string_Kofman} that these ``stable'' 
relics put an unavoidable lower bound on the inflationary scale in the
brane Universe. 
We will discuss this issue and the solution to the problem in
Sect.\ref{KK-string}. 
We will examine these competing conditions and show that it is possible
to construct realistic models of brane inflation using neither slow-roll
nor fine-tunings. It is worth re-emphasizing that the conventional
slow-roll approximations throughout this paper will not be used.

\section{Inflation in the brane Universe}
\label{Thermal-section}
\subsection{Thermal Inflation in the brane Universe}

\underline{Non-hybrid type}

Let us first explain the basic idea of the original thermal inflationary
model\cite{thermal_inf}.
In supersymmetric theories where the potential of the scalar
fields can be very flat, a field (flaton) can develop a large vacuum
expectation value $M$ even though its mass $m$ is very small.
The finite temperature in the early Universe can
hold such a field at zero, corresponding to a false vacuum
with energy density $V_0 \sim m^2 M^2$.
When the temperature falls below $V_0$, the thermal energy density
becomes negligible and an era of ``thermal inflation'' begins.
``Thermal inflation'' ends when the field rolls away from zero at a
temperature of order $m$, generating a number of e-foldings 
$N_e \sim \ln \left(\frac{T_i}{T_f}\right)$,
where $T_i$ and $T_f$ denote the temperature at the
beginning and at the end of thermal inflation, respectively.
In the conventional scenario of thermal inflation, where 
$T_i$ and $T_f$ are given by
$T_i  \sim   V_0^{1/4}$ and $T_f  \sim  m$,
the number of e-foldings is given by the two hierarchical scales $M$ and
$m$, 
\begin{equation}
\label{orig}
N_e \sim \frac{1}{2}\ln \left(\frac{M}{m}\right).
\end{equation}

The above idea of thermal inflation can be applied to the brane
inflationary models, since at high temperatures the points of the enhanced
gauge symmetries will be the local free energy minima,
 and the four-dimensional gauge dynamics can be reconstructed using
 brane dynamics.
In Ref. \cite{thermal_brane} Dvali argued that considering the coincident
branes corresponding to the enhanced gauge symmetry points one might
expect that branes are stabilized on top of each other.
For the observer living in the effective four-dimensional Universe, the
situation looks quite the same as the conventional thermal inflation
condition. 

\underline{Hybrid-type}

\begin{figure}[ht]
 \begin{center}
\begin{picture}(420,200)(0,0)
\resizebox{15cm}{!}{\includegraphics{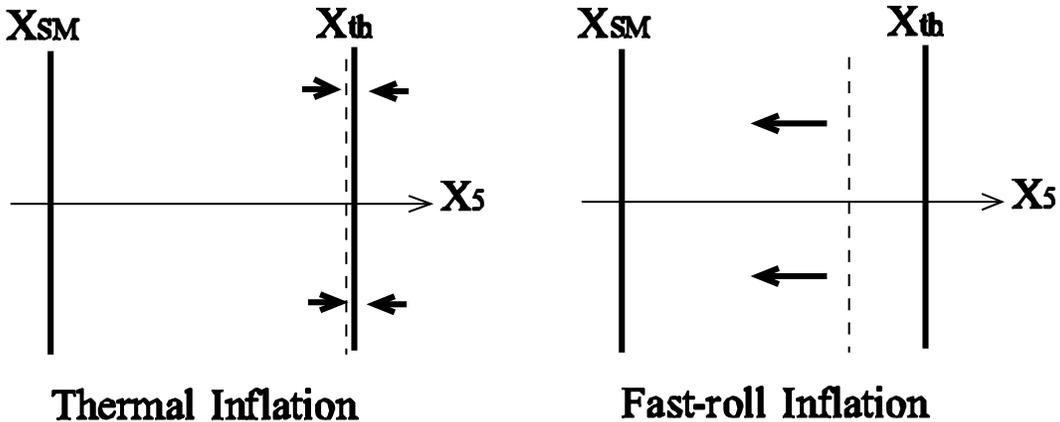}} 
\end{picture}
\caption{During thermal inflation, the moving brane is attracted to the
  fixed brane at the right-hand-side boundary. Then, at $T< T_f \sim m$,
the moving brane is detached from the right-hand-side boundary and begins
  to roll down to the anti-brane waiting at the left-hand-side boundary.}
\label{Fig:thermal}
 \end{center}
\end{figure}

A different approach in this direction is given in
Ref. \cite{matsuda_thermal_brane}, where thermal inflation is achieved 
with a hybrid-type potential.
The thermal inflationary models with the hybrid-type potential are first
discussed in the conventional four-dimensional theory, and then applied
to the brane Universe\cite{matsuda_thermal_brane}.
The most obvious advantage in using the hybrid potential is (1) the
enhanced number of e-foldings and (2) easy reheating. 
Since we are
considering the hybrid potential, we can use a potential in which the
vacuum energy during inflation does not depend on $m$.
Assuming that $V_0 \sim M^4$, we obtained the number of e-foldings
that is given by the two hierarchical scales $M$ and $m$;
\begin{equation}
N_e' \sim \ln \left(\frac{M}{m}\right).
\end{equation}
This result is twice as large as the original result given in
Eq. (\ref{orig}). 
Moreover, since we are considering the hybrid-type potential, the
requirement for the successful reheating does not put any critical bound
on $m$.
In the hybrid-type model, the trigger field that is much heavier than
the inflaton field induces reheating. As we will discuss later in this
paper, thermal inflation with the hybrid-type potential is more natural
in the brane Universe. We will use this idea to obtain the required
initial conditions in the new paradigm of brane inflation. We
schematically depict our basic idea in
fig.\ref{Fig:thermal}.\footnote{The ``trapping'' may be due to the 
non-trivial effect other than the thermal attraction.
For example, Kofman et.al. observed in  ref.\cite{beauty_is} 
that a scalar field trapped on a steep potential can induce a stage of
 universe acceleration, which is called ``trapped inflation''. }

\subsection{Brane inflation without slow-roll (thermal inflation)}
As discussed above, thermal brane inflation followed by fast-roll
inflation is very natural in the brane Universe. However, it is still
not clear if such an inflationary model can generate the required number
of e-foldings and curvature perturbations without producing unwanted
relics. In this section, we consider the inflationary expansion of the
model and calculate the number of e-foldings. A typical example is shown
in Fig. \ref{Fig:thermal}, where there is an
extra dimension denoted by $x_5$, and a 3-brane sitting at $x_5=X_{th}$
moving in this direction toward $x_5=X_{SM}$ due to a potential that can
be generated when supersymmetry is broken.
We consider two more branes fixed at $x_5=X_{th}$ and $x_5=X_{SM}$.
One may think that the fixed branes are the boundaries of the
$x_5$ direction, or they are sitting at the fixed points of a
moduli space.
The fixed branes may be 7-branes wrapping some four-cycle of the
compactified space other than $x_5$. 
We need to introduce at least one ``anti''-brane sitting at $x_5=X_{SM}$,
which is required to produce a hybrid-type inflation followed by brane
annihilation. 

It is assumed that initially the temperature is so high that the thermal
effects on the branes are so strong that they attract the moving brane
to the fixed brane at $x_5=X_{th}$.
As the temperature drops during thermal inflation, the attractive force
becomes weaker, and finally the force becomes so weak that 
the moving brane rolls down the
potential toward the ``anti''-brane sitting at $x_5=X_{SM}$.
Reheating is induced by the brane-antibrane annihilation at
$x_5=X_{SM}$.

To understand the characteristics of the model, we should first examine
the effect of the potential that induces the brane motion. 
Since we are
not considering slow-roll inflation, we must assume that the potential
is lifted at least by the $O(H_I)$ mass term,
$V_{eff}(\varphi) 
\simeq \frac{\alpha^2}{2} H^2 \varphi^2$. 
The generic form of the potential at a high temperature is given by
\begin{equation}
V(\varphi, T) = \pm\frac{1}{2} m^2(\varphi\pm \Delta)^2 + a T^2 \varphi^2
+ M^4,
\end{equation}
where $\varphi \propto x_5$ is a real scalar field which parameterizes the
position of the moving brane, and $a$ is a constant.
Here we set the field $\varphi$ such that thermal inflation occurs at
$\varphi=0$ and 
annihilation occurs at $\varphi=\varphi_{SM}$.
We introduced a new parameter $\Delta \ne 0$ because the minimum of the
potential might 
be shifted from either $\varphi=0$ or $\varphi_{SM}$.
We have assumed that the brane-antibrane attractive force is negligible
as far as the separation is significant. 
We will also assume (for
simplicity) that the potential is a monotonic function in the interval
$\varphi_{SM} + M <\varphi<0$.
Here $M$ is the typical mass scale of the inflating brane.
It is easy to calculate the minimum of the potential at a temperature
$T>m$, which is given by
\begin{equation}
\varphi_{min}(T)=\frac{- m^2 \Delta/2 }{\pm m^2/2 + a T^2}.
\end{equation}
The thermal excitation of the gauge field is significant as long as the
temperature is much higher than
$T_f\equiv|g\varphi_{min}|$, where $g$ is a gauge coupling constant.
On the other hand, at a temperature lower than $T_f$ there is no
significant thermal excitation, and thus the attractive force cannot
keep the moving brane at the right-hand-side boundary.
Here we would like to assume $g\sim 1$ for simplicity.
In the brane picture, the gauge field corresponds to the open string
stretched between branes.
Assuming a modest condition $m\ll \Delta \ll \varphi_{th}$, one can
find that thermal inflation is ended when the final temperature is
\begin{equation}
T_f \sim (m^2|\Delta|/a)^{1/3} \gg T_c,
\end{equation}
where $T_c=m/\sqrt{a}$ is the ``usual'' value of the final temperature.
It should be noted here that in the ``usual'' scenario there is an
undesired massless excitation at $T=T_c$, which induces an unwanted steep
spectrum.\footnote{See Ref. \cite{Thermal-fast-gong} for example. This is
the problem which can always appear in the so-called ``shoulder
inflation''. }
On the other hand, in our new scenario, the brane separation
$\varphi_{min}(T)$ becomes significant before such massless excitations
appear.
Therefore, the effective mass of the inflaton 
$m_{eff}^2 \equiv \pm m^2 + 2a T^2$ cannot become much smaller than
the Hubble constant $H \sim m$ even if the original mass term appears
with a negative sign.   
Moreover, unlike the non-hybrid type of the thermal inflationary model,
we can use a positive mass term for the field $\varphi$,
which cannot be canceled by the thermal mass term.
Introducing new constants $\alpha \equiv m/H$ and 
$\beta\equiv V_0^{1/4}/M$,
we obtained the number of e-foldings during thermal inflation as
\begin{eqnarray}
N_{e,1} &\simeq& \ln \left(\frac{T_i}{T_f}\right)\nonumber\\  
&\simeq &\frac{1}{3}\ln \left(\frac{M_p}{M}\right)
+\frac{1}{3}\ln \left(\frac{M_p}{\Delta}\right)+\frac{1}{3}\ln a
+\frac{1}{3}\ln \left(\alpha^2\beta \right).
\end{eqnarray}

\subsection{Brane inflation without slow-roll (Fast-roll inflation)}

After thermal inflation the motion of the moving brane is determined by
the $O(H)$ mass term.
Then, there will be another inflationary expansion called ``fast-roll''
inflation\cite{fast-roll}.
Since we will consider inflation in the KS throat in a later section, it
should be convenient to consider a case when the potential has a minimum
at $\varphi=\varphi_{SM}$ and the mass term appears with
a positive sign.
Then, it will be convenient to redefine the inflaton field $\varphi$ such
that the potential is given by 
\begin{equation}
 V(\varphi, \sigma) = \frac{1}{2}m^2 \varphi^2 
  + \frac{1}{2}\varphi^2 \sigma^2 +\frac{1}{4}\left(\sigma^2-M^2\right)^2,
\end{equation}
where $\sigma$ is the waterfall field.
The number of e-foldings elapsed during fast-roll inflation is
given by\cite{Hybrid-fast}
\begin{equation}
\label{shoulder_n}
N_{e,2} \simeq \frac{1}{F} \ln \left(\frac{\varphi_{th}}{M}\right),
\end{equation}
where $F$ is given by
\begin{equation}
F \equiv \frac{3}{2}\left(1-\frac{\sqrt{9-4\alpha^2}}{3}\right).
\end{equation}
Adding $N_{e,1}$ and $N_{e,2}$, we obtained the total number of
e-foldings
\begin{equation}
\label{Netotal}
N_e \simeq 
\frac{1}{3}\ln \left(\frac{M_p}{M}\right)
+\frac{1}{3}\ln \left(\frac{M_p}{\Delta}\right) +
\frac{1}{F} \ln \left(\frac{\phi_{th}}{M}\right) +\frac{1}{3}ln a
+\frac{1}{3}\ln \left(\alpha^2\beta \right)
.
\end{equation}
The required condition for the successful inflationary scenario
is\cite{book_EU}
\begin{equation}
\label{N-bound}
N_e +N_{pre}+N_{add} 
\ge N_{min}=53+\frac{2}{3}\ln\left(\frac{M}{10^{14}GeV}\right)
+\frac{1}{3}\ln \left(\frac{T_R}{10^{10}GeV}\right),
\end{equation}
where $N_{pre}$ and $N_{add}$ denotes the number of e-foldings elapsed
during a preceding(bulk) and an additional(weak) inflationary
expansions, if they existed.
From Eq. (\ref{Netotal}) and Eq. (\ref{N-bound}), it would be fair to
conclude that an inflationary scale higher than $10^{12}$GeV seems
unlikely in this model, if there is no fine-tunings nor
large number of $N_{pre}+N_{add}$.
From the above result, it is easy to understand that the total number of
e-foldings becomes larger as the inflationary scale $M$ becomes smaller.
Thus, in this model, in order to make a sufficient number of 
e-foldings, we must be careful about a lower bound for the inflationary
scale $M$. 
We will examine this problem in the following section.

\section{Generating the curvature perturbations at the end of brane
 inflation}
\label{Curva-Densi} 
Since the slow-roll potential is not the essential requirement in this
paper, it should be explained how one could obtain the required spectrum
of the curvature perturbations without using the usual mechanism. It is
well understood that adiabatic perturbations produced at later stages of
the fast-roll inflation have a much smaller amplitude (and a very steep
red spectrum), thus they cannot lead to desirable consequences. Besides
adiabatic perturbations, fast-roll inflation may produce isocurvature
fluctuations with a flat spectrum related to the perturbations of light
fields other than the inflaton field. Such perturbations may be
converted to adiabatic ones if the light field plays the role of the
curvaton, or it plays the essential role in the Lyth's mechanism of
generating the curvature perturbations at the end of inflation. In our
case, since it was assumed that thermal inflation occurs prior to
fast-roll inflation, the required mechanism for generating the adiabatic
perturbations must be consistent also with thermal inflation. Since the
vacuum energy is almost static during thermal inflation, thermal
inflation produces isocurvature perturbations with a flat spectrum
related to the perturbations of light fields if they (the light fields)
are protected by a symmetry and are ``not'' thermalized during thermal
inflation. 

Let us consider a specific example. Here we consider a light field
placed at the tip of the inflationary throat. The light field is
decoupled from the thermal plasma on the distant brane in the bulk, and
is very light due to the approximate isometries at the tip of the KS
throat. As we will discuss in Sect.\ref{KK-string}, the approximate
isometries will make the Kaluza-Klein modes with the angular momentum
dangerously long-lived. Thus, sooner or later, we must examine if the
stable Kaluza-Klein modes do not violate the conventional bounds for
unwanted stable relics. Our setups are quite generic in the brane
Universe. At first, there can be many $D_3$ and $\overline{D}_3$-branes
moving in the bulk.
Some of them may roll toward a $D7$-brane which wraps some
four-cycle in the bulk, and might be subsequently ``trapped'' for a
period of time\cite{beauty_is}.   
A mechanism of the ``trapping'' is discussed in Ref. \cite{beauty_is}.
Some of the trapped $D_3$ and $\overline{D}_3$-branes will annihilate
each other, reheating the Universe to make thermal inflation possible
at the $D_7$-brane.
It is also possible to consider a thermal inflationary model
where $N$($N>1$) $D_3$-branes are attracted to a fixed brane.

\subsection{The basic idea}
When one considers brane inflation, one may simply assume that the
typical inflaton field that determines the brane motion is the only
field that parameterizes the brane distance between the moving brane and
the waiting anti-brane. 
This idea is not completely incorrect, but requires
further consideration. For example, let us consider two inflaton fields
(i.e. two directions in the extra dimensions) $(\phi_1, \phi_2)$ with a
hybrid-type potential 
\begin{equation}
\label{3-1}
V(\phi_1, \phi_2, \sigma) =
\frac{m_1^2}{2}\phi_1^2 + \frac{m_2^2}{2}\phi_2^2
+\frac{\lambda}{2} \phi_r^2 \sigma^2
+\frac{1}{4}(\sigma^2-M^2)^2,
\end{equation}
where $\phi_i$ $(i=1,2)$ and $\sigma$ are taken to be real scalar
fields.
The above potential has global minima at 
\begin{equation}
(\phi_i, |\sigma|)=(0,M),
\end{equation}
and an unstable saddle point at 
\begin{equation}
(\phi_i, \sigma)=(0,0).
\end{equation}
Here $\phi_r\equiv \sqrt{\phi_1^2 + \phi_2^2}$ denotes the
brane-antibrane distance.\footnote{It would be useful to note here that
the usual ``brane distance'' is given by $r^2=\sum_i x_i^2$, where $x_i$
are the brane distances in the directions perpendicular to the inflating
branes.}
In a common case, there is no exact symmetry that protects all the
masses of the position moduli, and thus the equipotential surface of the
potential becomes an ellipsoid\cite{matsuda_elliptic}.
On the other hand, if there is an isometry in the compactified space, we 
can choose the field $\phi_2$ to parameterize the massless direction,
while a massive inflaton field $\phi_1$ determines the dynamics of the
brane inflation.\footnote{Unlike the conventional
models of ``successful brane inflationary models'', we do not assume any
isometry that protects the flatness of the inflaton $\phi_1$.
The field that is protected by an isometry is the secondary field
$\phi_2$, which is not the ``inflaton''.} 
As a consequence, we introduce a hierarchy between the mass scales;
$m_1 \gg m_2$. 
Inflation starts at large $\phi_1$ and ends at the point on the surface
$\phi_1^2 +\phi_2^2=M^2/\lambda$, where the
branes annihilate to reheat the Universe.

Since the secondary field $\phi_2$ is light during inflation,
the fluctuations along this direction is $\delta \phi_2 \simeq
H_I/2\pi$, where $H_I$ denotes the Hubble constant $H_I \sim M^2/M_p$
during inflation.

If the mass $m_1$ is not much heavier than $H_I$
and satisfies the condition 
$m_1 < \frac{3}{2}H_I$, the fast-roll inflation induced by the inflaton
field $\phi_1$ has the number of e-foldings
\begin{equation}
N = \frac{1}{F} \ln\left(\frac{\phi_{1,0}}
{\phi_{1,e}}\right)
\end{equation}
where $\phi_{1,0}$ and $\phi_{1,e}$ denotes the initial and the final 
value of the inflaton field $\phi_1$.
Although there are no sensible fluctuations related to the
perturbations of the inflaton field $\phi_1$, there is 
a generation of the curvature perturbations due to $\delta \phi_2$
at the end of inflation.
In the above formula for $N$, $\phi_2$-dependence appears in
$\phi_{1,e}= \sqrt{M^2-\phi_{2,e}^2}$.

The curvature perturbation $\zeta$ can now be calculated using the
method advocated in Ref. \cite{delta-N-Lyth}, which takes the form 
\begin{equation}
\label{toy1}
\zeta = \frac{\partial N}{\partial \phi_1}
 \frac{\partial \phi_1}{\partial \phi_2} \delta \phi_2
+ \frac{1}{2}\left\{
	      2 \frac{\partial^2 N}{\partial \phi_1^2}
	      \left(
\frac{\partial \phi_1}{\partial \phi_2}
\right)^2
	      + \frac{\partial N}{\partial \phi_1} 
	      \frac{\partial^2 \phi_1}{\partial \phi_2^2}
\right\}(\delta \phi_2)^2.
\end{equation}
It is amazing that this mechanism is still successful in generating the
desirable curvature perturbations even if the inflaton field $\phi_1$
does ``not'' satisfy the slow-roll conditions. 
Taking the mean value of the field $\phi_2$ as
 $\phi_2 \simeq \gamma M$ $(\gamma <1)$, we can calculate the curvature
 perturbations.
Assuming that the inflaton $\phi_1$ satisfies the fast-roll
condition\cite{fast-roll}, 
it is easy to obtain 
\begin{eqnarray}
\left.\frac{\partial N}{\partial \phi_1}\right|_e 
&=& -\frac{1}{F_1 \phi_{1e}} 
\nonumber\\
\left.\frac{\partial^2 N}{\partial \phi_1^2}\right|_e &=& \frac{1}{F_1 \phi_{1e}^2},
\end{eqnarray}
and
\begin{eqnarray}
\left.\frac{\partial \phi_1}{\partial \phi_2}\right|_e
&=&- \frac{\phi_{2e}}{\phi_{1e}}\nonumber\\
\left.\frac{\partial^2 \phi_{1}}{\partial \phi_{2}^2}\right|_e
&=&-\frac{1}{\phi_{1e}}\left(\frac{\phi_{2e}^2}{\phi_{1e}^2}+1\right).
\end{eqnarray}
As a result, we obtained $\zeta_\Delta(x)$ which is given by
\begin{equation}
\label{zeta1}
\zeta_\Delta(x) = \frac{\gamma H_I}{2\pi F_1 (1-\gamma^2)M}
+\frac{(2\gamma^2+1 )H_I^2}{8\pi^2F_1 (1-\gamma^2)^2 M^2}.
\end{equation}

\begin{figure}[ht]
 \begin{center}
\begin{picture}(420,220)(0,0)
 \resizebox{12cm}{!}{\includegraphics{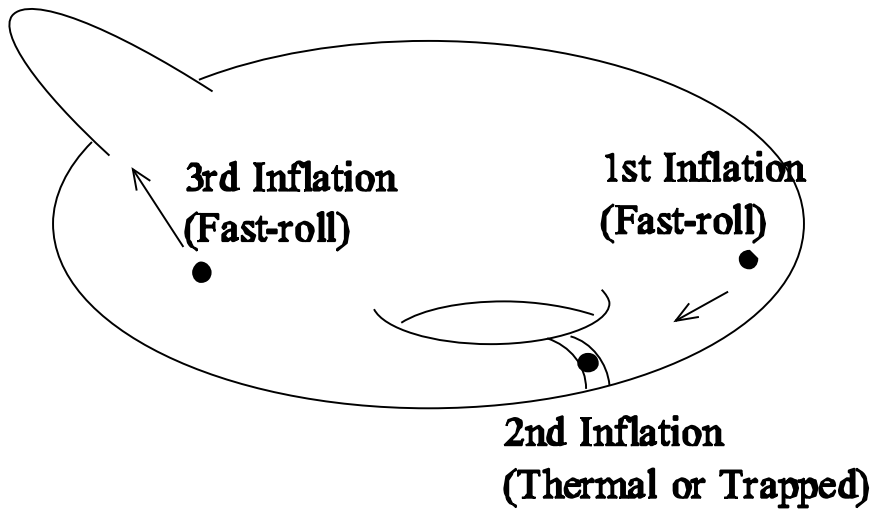}} 
\end{picture}
\caption{A typical situation in the brane Universe}
\label{Fig:multi}
 \end{center}
\end{figure}

As far as one is considering the single-shot inflation in the brane
Universe, the typical scale of the Hubble constant during inflation is
given by $H_I \simeq M^2/M_p$, suggesting that the above result puts a
strict bound on the inflationary scale $M/M_p \sim 10^{-5}$,
if there is no fine-tunings.
However, it is always possible to consider an additional stage of
inflation in the bulk (or in a different throat), which can produce
another inflationary scale $H_I'\gg H_I$.
For example, let us consider a case where initially many $D_3$ and
$\overline{D}_3$-branes are moving in the bulk.
Some of them rolls toward a fixed $D_7$-brane in the bulk, where
they will be ``trapped'' for a period of time\cite{beauty_is}. 
Some of the trapped $D_3$ and $\overline{D}_3$-branes will annihilate
each other, reheating the Universe to make thermal inflation possible.
During this period, the light field $\phi_2$ at the bottom of a throat
is protected by an approximate isometry at the tip of the KS throat
where the final target-brane lives.
If so, the fluctuation related to the light field $\phi_2$ is given by
$\delta \phi_2 \simeq H_I'/2\pi$, which is much larger than $H_I/2\pi$.
See Fig.\ref{Fig:multi} to understand a typical situation.
The strict bound $M/M_p \sim 10^{-5}$ obtained above can be removed if
there is an additional period of inflation, which has an energy scale
higher than the energy scale of the last inflation.
The lowest scale of the last inflation would be realized when it occurs
in the standard-model throat, where the inflationary scale
$M$ is as low as O(TeV).
In this extreme case, the inflationary scale in the bulk could be as low
as $M' \sim 10^{8}$GeV.

\section{Comment on Dangerous relics and baryogenesis}
\label{KK-string}
\subsection{Dangerous relics}
As we have discussed above, this new scenario favors a low inflationary
scale since a large number of e-foldings requires a low inflationary
scale in the order of 
$M < 10^{12}$GeV, if there are no fine-tunings. 
Although (naively) the mechanism
for generating the curvature perturbations at the end of inflation
cannot produce enough perturbations if the inflationary scale is lower
than
$10^{13}$GeV, this problem can be solved by introducing a preceding
stage of inflation in the bulk, which has the Hubble constant
$H_I' \sim 10^{-5} M$.
If so, fast-roll brane inflation in the KS throat may occur in the
standard-model throat, generating a large enough number of e-foldings
provided that a preceding inflationary stage has the inflationary scale
$M' \sim 10^{8}$GeV.

The question as to whether there are constraints coming from other
cosmological considerations requires an answer. Do they disrupt the
above scenarios? It is assumed the reader is knowledgeable in the
various constraints in this direction, however the constraints are
sometimes based on some specific conditions that are highly
model-dependent. Thus, it would be helpful to start with historical
arguments. 
Let us first consider a generic condition that places a significant
bound on the scale of the usual hybrid-type inflationary model. During
hybrid inflation, the inflaton field should have a significant coupling
to the trigger field. As discussed in Ref. \cite{TeV-hybrid},
this coupling may induce a one-loop contribution to the inflaton
potential, which may then ruin the slow-roll approximation and the
generation of the structure of the Universe. Therefore, as far as one is
considering a scenario where the slow-roll approximation plays a
significant role, one cannot have a successful model of hybrid inflation
with a scale lower than $10^9$ GeV.
It might be thought that this bound contradicts the above argument, and
start to suspect that the last inflation cannot occur in the
standard-model throat. However, in the analysis of
Ref. \cite{TeV-hybrid}, the crucial condition came from the requirement
of generating the curvature perturbations from the
``slow-roll'' inflaton field.
Thus, our model does not suffer from this condition. One may think that
the TeV-scale inflation could be successful if the curvaton plays its
role. However, as we have discussed in
Sec.1, there is a strict bound on the inflation scale even if the
curvaton is introduced\cite{Lyth_constraint}.
This serious bound could be avoided by introducing additional inflation
\cite{matsuda_curvaton}, however this mechanism would require many
non-trivial assumptions in order to make the additional phase
transitions viable in this scenario. Besides the above conditions, other
conditions coming from the fact that the typical mass scale of the brane
bound the maximum value of a scalar field on the brane must be
considered; in models with a low fundamental scale
$M \sim$O(TeV), the inflation field must be a bulk field if it ``rolls
in'' from a distance. In this case, one cannot exclude the serious
constraint coming from the decaying Kaluza-Klein mode\cite{matsuda_KK},
since such bulk fields always couple to the Kaluza-Klein states.
The constraint appears even if one discards the generation of the
curvature perturbations due to the ``slow-roll'' inflaton field.
Recently, it has been suggested in Ref. \cite{KK-string_Kofman} that 
the constraint coming from the Kaluza-Klein mode becomes much more
serious in the brane inflationary models with some isometries in the
compactified space. In the brane Universe, the stability of the
Kaluza-Klein mode with the angular momentum related to the isometry
depends on the scale of the inflation scale $M$, giving 
the stringent condition $M >10^{12}$GeV\cite{KK-string_Kofman} for the
``safe-decay''.
Of course, the above condition is quite significant in our scenario and
seems to exclude low-scale inflation in the standard-model throat. More
recently, however, it has been discussed in Ref. \cite{warped-reaheating}
that studying a more detailed thermodynamic 
evolution of the heating process, especially that of the KK particles,
many qualitatively different results compared to the original result
can be found. 
The new result obtained in Ref. \cite{warped-reaheating} is quite helpful
in reducing the scale of inflation, since it allows inflation in the
standard-model throat\footnote{A conserved
angular momentum may be significant\cite{Lefteris, Grandi}}.

One may also worry about dangerous cosmological defects that may be
produced during brane annihilation. Although cosmic strings can put an
upper bound on the inflationary scale of the last
inflation\cite{string_relic},
 it is looser than the above bound obtained in the calculation of the
 number of e-foldings.
Besides the cosmic strings, monopoles and walls may be produced as a
consequence of brane creation or brane deformation that may occur during
or after the reheating epoch\cite{matsuda_brane_monopolesandwalls}. 
It is known that these defects might put a strict upper bound on the
inflationary scale\cite{matsuda_brane_relics},
if there is a non-trivial structure in the compactified space. 
A natural solution to the domain wall problem in a typical supergravity
model is discussed in Ref. \cite{matsuda_wall}, where the required
magnitude of the gap in the quasi-degenerated vacua is induced by $W_0$
in the superpotential.
The mechanism discussed in Ref. \cite{matsuda_wall} is natural since the
constant term $W_0$ in the superpotential is necessary so as to cancel
 the cosmological constant. 
The bound obtained in Ref. \cite{matsuda_brane_relics} is severe, but
it is also known that the structure that is required
to produce the dangerous defect configuration of the brane does not
appear in the known example of the KS throat. 
Of course, it is not clearly understood whether it is possible to obtain
the complete Standard Model(SM) in the known example of the KS
throat. Thus, it should be fair to conclude here that this problem is
unsolved and requires further investigation together with the
construction of the complete set of the SM model in the brane Universe. 

\subsection{Baryogenesis}
In lowering the inflationary scale it might be thought that it would be
difficult in obtaining enough baryon number asymmetry of the Universe
(BAU), since the requirement of the proton stability puts a strict bound
on the baryon-number-violating interactions. This speculation is indeed
true. The old scenario of the baryogenesis with a decaying heavy
particle cannot work if the fundamental scale becomes as low as the O
(TeV) scale\cite{low-x-decay-baryo}.
One can solve this problem by introducing cosmological defects that
enhance the breaking of the baryon number conservation in the core
\cite{low-x-decay-baryo_matsuda}.
There is also a problem in the scenario of Affleck-Dine
baryogenesis\cite{AD}, since the expectation value of a field on a brane
cannot become much larger than the typical mass scale of the
brane\cite{low-AD-problem}. 
One can solve this problem by introducing non-trivial defect
configuration structures\cite{low-AD-solution_matsuda}.
Moreover, in Ref. \cite{low_baryo}, the reader will find more arguments
about the mechanism of baryogenesis with low-scale inflationary scale.
Although there is no ultimate solution to this problem,
it is not incorrect to expect that baryogenesis could be successful even
if the inflationary scale is as low as the TeV scale.

\section{Conclusions and Discussions}
\hspace*{\parindent}
We studied a typical situation in brane inflationary models without
using the conventional slow-roll approximations. 
Based on the sensible idea advocated recently in Ref. \cite{delta-N-Lyth},
we have found a new brane inflation paradigm.
 We examined the conditions
for constructing a large enough number of e-foldings and for generating
the curvature perturbations. We also examined whether dangerous unwanted
relics could be produced. Benefits of our scenario are obvious: we do
not have to be concerned about the famous
$\eta$-problem if an isometry appears in the structure of the
compactified space.
The most useful isometry example is the one appearing at the KS throat,
which is, of course, a natural candidate for the ``realistic'' brane
Universe. 

\section{Acknowledgment}
We wish to thank K.Shima for encouragement, and our colleagues at
Tokyo University for their kind hospitality.
\appendix
\section{More on the junctions}
Our scenario is based on the simple idea that 
(1) thermal brane inflation is generically supposed to occur after
primary inflation and (2) there could be some inflationary stage after
thermal brane inflation, which is sometimes called by the name ``shoulder
inflation''. 
Hence generically one should consider three different kinds of inflation
if the inflationary model contains thermal inflation.
However, such shoulder inflation has been supposed to be very short or
simply disregarded, because of the serious problem related to the
overproduction of PBHs.
The PBH overproduction puts a severe bound on models with conventional 
shoulder inflation\cite{Thermal-fast-gong}.
One of the important advantages in our scenario is that this condition
does not apply to brane inflation models in which the potential is not
the rigorous function of the brane distance.\footnote{A similar
situation has been considered in Ref. \cite{beauty_is}.}
Of course, if the broken isometry of the internal space is simply due to
the existence of the other brane, the potential that lifts the flat
direction related to the isometry must be a rigorous function of the
brane distance.
On the other hand, if the isometry is explicitly broken due to the
non-trivial structure of the compactified space, for example by the
existence of fluxes or warping, the potential for the brane position
could be a function determined by the structure of the compactified
space.
In this case, the potential for the brane position is not a rigorous
function of the brane distance. 
Using  a simple inflationary model in which thermal brane
inflation occurs at a boundary (or at a fixed brane) in the $x_5$
direction, we showed that the serious bound related to the
unwanted PBH production can be avoided if the potential for the
moving brane is not the rigorous function of the brane distance.
The boundary in the $x_5$
direction is the thermal attractor during thermal inflation, and a
``shift'' from the thermal attractor grows during thermal 
inflation.
This ``shift'' induces the desired mass for the unwanted massless mode,
and reduces the overproduction of unwanted PBHs.
It would be important to note again that in our analysis the potential is
not the rigorous function of the brane distance.
This is what happens in generic models with warped internal space, or
moduli stabilization by the fluxes\cite{KKLT}.
The idea that the potential for the brane position is not the rigorous
function of the brane distance plays the most important role in our
analysis of generating 
the curvature perturbation at the end of inflation.
In our present model the potential for $\phi_1$ and $\phi_2$ should not
be the rigorous function of the brane distance that is given by $\phi_r$.
The difference between $m_1$ and $m_2$ cannot be justified if the
potential is the rigorous function of the brane distance.
In order to make $m_1 \gg m_2$, we have assumed that the light field
$\phi_2$ is related to some enhanced angular isometry of the compactified
space.
This is a typical situation in the KS-throat where angular isometries
are enhanced at the tip of the throat\cite{matsuda_elliptic,
Riotto-Lyth}\footnote{Of course there could be 
many other cases in which some angular isometries remain in realistic
compactifications. We do not exclude other models with angular
isometries.} 
and  what happens in the potential Eq. (\ref{3-1}) for large
$\phi_r$.
On the other hand, the end-line of the hybrid
inflation is determined by the brane distance $\phi_r$ if the
brane-brane interaction triggers the end of inflation.
We think some comments are required to explain the origin of the
enhanced isometries in the KS throat.
Consider the dS background of KKLT\cite{KKLT}.
The idea is that the stabilization of the K\"ahler modulus leads to a
vacuum with a negative cosmological constant, however the vacuum can be
lifted by adding an $\overline{D3}$-brane that sits at the tip of the
throat. 
Introducing an $\overline{D3}$-brane to the warped background one will
find new moduli that corresponds to the position of the
$\overline{D3}$-brane on the compact space.
The potential for the moduli in the KKLT background is not trivial.
The $\overline{D3}$-brane is not free to move in the throat direction
since they have a potential proportional to the warp factor which
stabilizes the $\overline{D3}$-brane at the tip of the throat.
Although the throat direction is stabilized by the potential,
the $\overline{D3}$-brane can still move on the $S^3$ at the tip of the
throat. 
In the exact KS solution, $S^3$  at the tip of the throat is exact.
The symmetry of $S^3$ is broken by placing the $\overline{D3}$-brane as
$SO(4)\rightarrow SO(3)$, which gives rise to three massless moduli.
The moduli correspond to the three coordinates of the position of the
$\overline{D3}$-brane on the $S^3$.
Besides the corrections in the IR-boundary which is expected to be
small, there would be another correction from the UV-boundary.
Generically the UV-boundary is supposed to be the Calabi-Yau manifold
where the background 
deviates from the original KS solution away from the tip.
The corrections from the UV-boundary may explicitly break the $SO(4)$
symmetry generating masses for the moduli.
Although the deformation of the theory 
generates $O(H)$ mass for the moduli at the UV-side, the contribution
is exponentially smaller than the typical mass scale at the tip.
Therefore, in a brane inflationary scenario where the
$\overline{D3}$-brane is fixed at the tip of the throat while the moving
brane comes from the root, the fluctuation of the $\overline{D3}$-brane
in the direction of $S^3$ is important although the fluctuation is
negligible for the moving brane at the UV-side.
The target brane fluctuates in the direction of $S^3$ but the moving
brane does not fluctuate.
Angular isometry may appear in other string models, which may or may not
protect the mass of the light field $\phi_2$.

Here we would like to explain more for the three inflationary
stages; primary inflation, thermal inflation and  shoulder inflation.
In order to induce thermal brane inflation after the primary
inflation, there should be a moving brane that feels the thermal
attractive force from the fixed brane(or the boundary).
The required situation is also explained in the original
paper\cite{thermal_brane}, however one might think the conditions are 
 artificial.
We think the most artificial point is the existence of a moving brane,
which is supposed to appear not at the stable point of the potential but
at the unstable point near the boundary, and then sticks to the boundary
due to the thermal attractive force. 
Of course it is possible to assume that the boundary is the
quasi-stable point during the primary inflation, since there could be 
corrections of $O(H)$ during primary inflation which might give
 large positive mass to the open string mode.\footnote{Alternatively, 
one may consider 
``trapped inflation''\cite{beauty_is} instead of thermal brane
inflation.} 
This is the common ambiguity that exists whenever one considers thermal
brane inflation.
This is the weak point of our argument.

Another thing that we believe to be explained more is the origin of the
large-scale fluctuation of $\phi_2$.
One might think that $\delta \phi_2$ must be generated during the last
(shoulder) inflationary period since the curvature perturbation is
generated at the end of this period.
This is not the case in Lyth's scenario for generating the curvature
perturbation at the end of inflation\cite{delta-N-Lyth}.
Since the curvature perturbation depends on the fluctuation $\delta N$,
where $N$ is the total number of e-foldings elapsed after horizon exit,
 the large-scale fluctuation of the light field $\phi_2$ must be
 generated at the initial few e-foldings of $N$.
Hence, the required large-scale fluctuation of 
the light field $\phi_2$ should be generated during the
preceding inflationary periods, if the last (shoulder) inflationary
expansion is not enough.

We would like to consider the natural range of the number of e-foldings
related to the last (shoulder) inflation.
If one excludes ``small'' mass $m < 0.1 H$, the upper limit for the
number of the e-foldings elapsed during shoulder inflation is given
by Eq.(\ref{shoulder_n}), where $F\simeq 0.003$ for $m =0.1 H$.
This ``minimum'' value for the inflaton mass $m\simeq 0.1H$ leads to
the ``maximum'' value for the number of the e-foldings $N_{max} >100$, 
which is obviously large. 
On the other hand, for the inflaton mass $m = H$ the factor
$F$ is enhanced up to $F\simeq 0.38$ and it seems unlikely to have $N=60$
without considering additional thermal or primary inflation.

\end{document}